\documentclass[a4paper,11pt]{revtex4}

\usepackage{graphicx}
\usepackage{amsmath}
\usepackage{amsfonts}
\usepackage{amssymb}
\usepackage{bm}
\usepackage{epsfig}
\usepackage{graphicx}

\usepackage[top=1.0cm, bottom=1.0cm, left=2.0cm, right=1.5cm]{geometry}

\begin{document}
\title{THE GAP VS CRITICAL TEMPERATURE RATIO IN PEIERLS-TYPE PHASE TRANSITIONS}
\author{\underline{G.Y. Chitov}}
\affiliation{Department of Physics, Laurentian University, Sudbury, ON, P3E 2C6
Canada\\
\centerline{email: gchitov@laurentian.ca}}

\begin{abstract}
We analyze a 2D spin-pseudospin model, where the pseudospins represents the charge
degrees of freedom. The model is known to undergo a phase transition with the simultaneous
appearance of the long-range charge order and the spin gap. We show how the gap vs critical
temperature ratio (also called the BCS ratio) gets renormalized from the classical
non-interacting value. This value is also universal in the sense that is does not depend on the
microscopic parameters of the model, and must be the same for various types of the Peierls-like
transitions where the spin gap is accompanied by the structural, orbital or charge order.
\end{abstract}

\maketitle

$\bullet$ {\em \underline{Introduction.}} --
There is a large number of materials which demonstrate phase transitions
into states with a spin gap accompanied by some types of the structural, charge
or orbital order. The most canonical example is the spin-Peierls transition
\cite{Pouget01,CF79}. The interplay of charge, spin and orbital degrees of freedom is
known to produce some exotic phases in transition metal oxides \cite{Kugel82,Oles10}.
The dimerized Peierls states driven by the superstructures of the orbital order are
reported in some spinels \cite{Khomskii05}. Our main motivation comes from the recent
work on the spin-SAF transition in the quarter-filled ladder compound $\rm NaV_2O_5$
\cite{Most,GC04,GCStack04,CGEPL05,GCFNT05}, where the Super-Anti-Ferroelectric (SAF)
long-range charge order occurs together with the spin gap.
Several other layered vanadate compounds demonstrate transitions when the spin
gaps occur simultaneously with charge ordering. In particular, the spin-SAF
transition was recently reported in Zn(pyz)$\mathrm{V_4O_{10}}$ \cite{Yan07}.

If, in all seemingly different Peirels-like states the spin gap is induced by the
dimerization due to structural, orbital or charge long-range order, then there should
be some universal parameters unifying all such transitions. A good candidate is the BCS
ratio we calculate here for the case of the spin-SAF transition. It does not depend on
the model microscopic parameters and matches the value found earlier for the spin-Peierls
transition.

\begin{figure}[h]
\epsfig{file=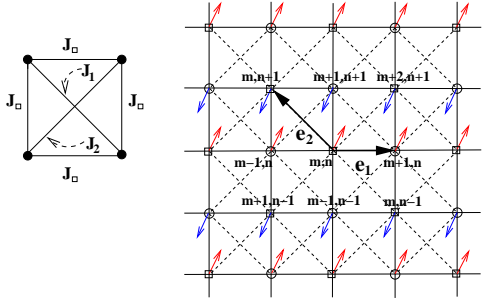,width=0.35\textwidth,angle=0}
\caption{Couplings on an elementary plaquette of the 2D nn and nnn Ising model
(left) and an example of the 4-fold degenerate SAF order (right). The SAF is
also called \textit{columnar} or \textit{stripe} order in some recent
literature. The degeneracy of the SAF state is due to the $Z_2 \otimes Z_2$
symmetry with respect to the spins of each of the sublattices (labeled by
circles or squares) being flipped. The lattice vectors are parameterized by the
integers $(m,n)$ in the skewed basis $\mathbf{r} = m \mathbf{e}_1 +n
\mathbf{e}_2$. } \label{PlaqSAF}
\end{figure}

%
%
$\bullet$ {\em \underline{Spin-Pseudospin Model and Analysis.}} --
We analyze here a spin-pseudospin model which consists of the Ising pseudospins $\bm{\mathcal
T}$ coupled to the Heisenberg  spins $\mathbf{S}$ which reside on the same sites of a square
lattice. The pseudospin sector is given by the Ising Model in a Transverse Field (IMTF).
An elementary plaquette and the couplings in this model are shown in Fig.~\ref{PlaqSAF}.
The Hamiltonian of this IMTF reads:
\begin{equation}
\label{HIMTF}
    H_{\textrm{IMTF}}= \frac12 \sum_{nn,nnn} J_{\sharp}
    \mathcal T^x_{\mathbf{k}} \mathcal T^z_{\mathbf{l}}
    - \Omega \sum_{\mathbf{k}} \mathcal T^z_{\mathbf{k}}~.
\end{equation}
The spin sector is given by the Heisenberg chains parallel to the $J_1$ diagonals
$H_S= J \sum_{m,n} {\textbf S}_{mn}{\textbf S}_{m,n+1}$.
The spin-pseudospin coupling is:
\begin{equation}
\label{HST}
    H_{ST}=   \frac12 \varepsilon  \sum_{m,n} {\textbf S}_{mn}{\textbf S}_{m,n+1}
    \big( \mathcal T^x_{m+1,n+1}- \mathcal T^x_{m-1,n}\big)~,
\end{equation}
with the total Hamiltonian of the model:
\begin{equation}
\label{Htot}
    H= H_{\textrm{IMTF}}+ H_S+ H_{ST}~.
\end{equation}
The spin and pseudospin operators satisfy the standard $SU(2)$ algebra,
while $S$ and $\mathcal T$ commute.

The spin-pseudospin model (\ref{Htot}) and some of its modifications were
proposed and analyzed in the earlier related work
\cite{GC04,GCStack04,CGEPL05,GCFNT05} in the context of the quarter-filled
ladder compound $\rm NaV_2O_5$, where the pseudospins correspond to the charge
degrees of freedom. For the physically interesting couplings \cite{GCFNT05} the
classical Ising model ($\Omega=0$) orders into the 4-fold degenerate
Super-Anti-Ferromagnetic (SAF) phase \cite{FanWu69} shown in
Fig.~\ref{PlaqSAF}. The SAF pattern appears in various contexts, and it is also
called \textit{columnar} or \textit{stripe} order in some more recent
literature. The SAF charge order occurs along with dimerization and gap in the
spin sector \cite{GCFNT05}. That is what we called the spin-SAF transition
\cite{GC04,GCStack04,CGEPL05,GCFNT05}. Note that the spin gap is due to the
frozen phonon displacements at the spin-Peierls transition, while the charge
plays the role of phonons at the spin-SAF transition.

The molecular-field approximation is applied for the pseudospins \cite{GC04},
while the spin sector is treated via minimization of the exact free energy of
the dimerized Heisenberg $XXX$-chain
\begin{equation}
\label{HXXX}
    H_{\textrm{xxx}}=\sum_{n}
     J(1+(-1)^n \delta) {\textbf S}_{n}{\textbf S}_{n+1}~.
\end{equation}
The specific free energy of the spin chain $f_s(T, \delta)$ is an analytic
function at $T \neq 0$ and can be expanded over $\delta$ as $f_s(T,\delta) =
f_s(T,0) - \frac{J}{2} \eta(T,0) \delta^2 + \mathcal O(\delta^4)$, where
$\eta(T,0)$ is called the static dimerization susceptibility.

%
%
$\bullet$ {\em \underline{Spin-SAF phase transition. XY Spin Chain:~Free Fermions.}} --
It is straightforward  to obtain in a closed form the specific free
energy of the XY spin chain mapped onto the spinless non-interacting Jordan-Wigner
fermions. To leading order \cite{GC04}
\begin{equation}
\label{etaxy0}
    \eta^{\textrm{\tiny XY}}(T,0)=\frac{1}{\pi} \ln \frac{\mathbb C J}{T}
    + \mathcal O \Big( \frac{T^2}{J^2} \Big)~,
\end{equation}
where $\mathbb C \equiv \frac{4}{\pi \textrm{e}^{1- \gamma}} $,
$\gamma=0.5772...$ is Euler's constant. In the region $\mathcal{J} < \mathcal{J}_c$
(where $\mathcal{J}_c=2 \Omega$ is the mean-field value of the QCP of the IMTF (\ref{HIMTF})
\cite{GCFNT05}), the critical temperature is given by the BCS-type solution
\begin{equation}
\label{TcBCS}
 T_c \approx \mathbb C J  \textrm{exp}
 \Big[- \frac{\pi J}{ \varepsilon^2}(\mathcal{J}_c -\mathcal{J} ) \Big]~.
\end{equation}
The ground-state dimerization is
\begin{equation}
\label{DimBCS}
 \delta \approx  \frac{4}{\mathrm{e}}\textrm{exp}
 \Big[- \frac{\pi J}{ \varepsilon^2}( \mathcal{J}_c -\mathcal{J}) \Big],~~T=0~.
\end{equation}
In the case of free fermions the spin gap depends linearly on dimerization,
$\Delta^{\textrm{\tiny XY}}=J \delta$. So the ratio of the zero-temperature
spin gap ($\Delta_\circ$) and the critical temperature (a.k.a the BCS ratio) in
the regime $\mathcal{J}_c <\mathcal{J}$ is
\begin{equation}
\label{BCSratioXY}
    \frac{\Delta_{\circ}^{\textrm{\tiny XY}}}{T_c}=
    \frac{4}{\mathrm{e} \mathbb{C}}=\frac{\pi}{\mathrm{e}^{\gamma}}
= 1.76...~,
\end{equation}
which coincides exactly with the classical result for the superconducting
gap in the BCS theory \cite{AGD}.

%
%
$\bullet$ {\em \underline{Spin-SAF phase transition. XXX Spin Chain:~Interacting Fermions --
sine-Gordon Model.}} --
The Heisenberg spin chain can be mapped onto the model of interacting Jordan-Wigner spinless
fermions, and the low-energy sector of the fermionic Hamiltonian in its turn can be bosonized \cite{Tsvelik}.
Neglecting the marginal term, the dimerized spin chain maps onto the
sine-Gordon model \cite{Tsvelik}
\begin{equation}
\label{HsG}
    v^{-1}H_{\mathrm{sG}}=\frac12 \int dx \big(\Pi^2+(\partial_x\phi)^2 \big)
    +2 \mu \int dx \cos \sqrt{2 \pi} \phi~,
\end{equation}
where $v=\frac{\pi}{2}J$ is the bosonic velocity and $\mu = \frac{A_\epsilon}{\pi} \delta$.
The relevant perturbation of the free bosonic part of the Hamiltonian (\ref{HsG}) comes from
the spin dimerization term $(-1)^n {\textbf S}_{n}{\textbf S}_{n+1} \sim A_\epsilon \cos \sqrt{2 \pi} \phi$.
The amplitude $A_\epsilon$ is not known exactly yet, but
according to the approximate calculations of Orignac \cite{Orignac04}
\begin{equation}
\label{Ae}
    A_\epsilon = \frac{3}{\pi^2} \Big(\frac{\pi}{2} \Big)^\frac14
\end{equation}
Using the sine-Gordon model (\ref{HsG}) to approximate the low-energy sector
of the dimerized Heisenberg chain (\ref{HXXX}), the free energy of the latter
reads to leading order \cite{Tsvelik,OrignacChitra04}
\begin{equation}
\label{fs}
    f_s(T, \delta)=-J t_\circ - \frac13 \frac{T^2}{ J}
    - \frac12 \frac{J^2 a_\circ}{T} \delta^2 ~,~~ \textrm{where}~~
    a_\circ \equiv \frac14 \bigg( \frac{\Gamma(1/4)}{\Gamma(3/4)} \bigg)^2 A_\epsilon^2
    \equiv \vartheta_1 A_\epsilon^2~,
\end{equation}
and $t_\circ \equiv \ln 2 -\frac14 = 0.4431...$. Then the dimerization
susceptibility to lowest order
\begin{equation}
\label{etaC}
    \eta(T,0)=\frac{a_\circ J}{T}~.
\end{equation}
This function was first calculated by Cross and Fisher \cite{CF79} for their spin-Peierls
transition theory. The amplitude $A_\epsilon$ (\ref{Ae}) suggested by Orignac \cite{Orignac04}
gives $a_\circ=0.253418$, very close to the original bosonization result
$a_\circ^{\textrm{\tiny CF}} \approx 0.26$ \cite{CF79}.

\begin{figure}[h]
\epsfig{file=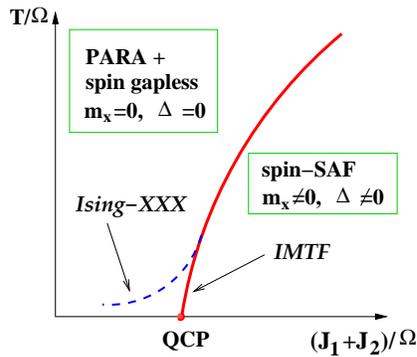,width=0.30\textwidth,angle=0}
\caption{Phase diagram of the coupled Ising-XXX model. Axes are not in scale.
The solid red line shows the critical temperature $T_c$ separating the
disordered and SAF phases of the decoupled IMTF. Its QCP corresponds to
$\mathcal{J}_c= 2 \Omega$  in the mean-field approximation. $(\mathcal{J}
\equiv J_1+J_2)$. The dashed blue line shows $T_c$ separating the disordered +
spin gapless phase from the spin-SAF phase of the coupled Ising-XXX model.} \label{PhDiag}
\end{figure}

The interactions of the Jordan-Wigner spinless fermions of the XXX chain  modify considerably the
critical properties of the Ising-XXX model (\ref{Htot}). The inverse dependence $\eta \propto T^{-1}$
(\ref{etaC}) results in the power-law dependence of $T_c$ on couplings:
\begin{equation}
\label{TcXXX}
T_c =\vartheta_1 A_\epsilon^2
\frac{ \varepsilon^2}{\mathcal{J}_c-  \mathcal{J}}~,~~\mathcal{J} < \mathcal{J}_c.
\end{equation}
The phase diagram of the Ising-XXX model is shown in Fig.~\ref{PhDiag}.

From the ground-state specific energy of the sine-Gordon model \cite{Baxter,AlZam95}
\begin{equation}
\label{f0sg}
    f_{\mathrm{sG}}(0,\mu)=-v \frac{M^2}{4} \tan \frac{\pi}{6}~,~ \textrm{where}~~
     M= \frac{2}{\sqrt{\pi}} \bigg(\pi \mu \frac{\Gamma(3/4)}{\Gamma(1/4)} \bigg)^{2/3}
    \frac{\Gamma(1/6)}{\Gamma(2/3)}
\end{equation}
($M$ is the dimensionless soliton mass) we get the ground-state energy of the
Heisenberg chain (\ref{HXXX}):
\begin{equation}
\label{f0s}
    f_s(0, \Delta_\circ)=-J t_\circ
    -\frac{1}{2 \pi \sqrt{3}}\frac{\Delta_\circ^2}{J} ~,
\end{equation}
where the zero-temperature spin gap $\Delta_\circ= \frac{\pi}{2} J M$ is related to the dimerization
as follows:
\begin{equation}
\label{Delta0}
    \Delta_\circ= J \sqrt{\pi} \bigg( \frac{\Gamma(3/4)}{\Gamma(1/4)} \bigg)^{2/3}
    \frac{\Gamma(1/6)}{\Gamma(2/3)} A_\epsilon^{2/3} \delta^{2/3} ~.
\end{equation}
The ground-state spin gap:
\begin{equation}
\label{Delta0sg}
 \Delta_\circ =  \vartheta_2 A_\epsilon^2
 \frac{ \varepsilon^2}{\mathcal{J}_c -\mathcal{J}}~,~~\textrm{where}~~
 \vartheta_2 \equiv \frac23 \sqrt{\frac{\pi}{3}}
    \bigg( \frac{\Gamma(3/4)}{\Gamma(1/4)} \bigg)^2
    \bigg( \frac{\Gamma(1/6)}{\Gamma(2/3)} \bigg)^3
 \end{equation}
Combining Eqs.~(\ref{TcXXX},\ref{Delta0sg}) we obtain the BCS ratio:
\begin{equation}
\label{BCSXXX}
    \frac{\Delta_\circ}{T_c}=\frac{\vartheta_2}{\vartheta_1}=
    6 \sqrt{3} \frac{\big(\Gamma(1/3) \big)^9}{\big( \Gamma(1/4) \big)^8}
    =2.47...
\end{equation}
The same BCS ratio for the spin-Peierls transition was first obtained by
Orignac and Chitra \cite{OrignacChitra04}. We would like to stress that the
above result is exact for the IMTF coupled to the sine-Gordon model, i.e., when
the marginal terms in the Heisenberg spin Hamiltonian are neglected.
Similar to the non-interacting result (\ref{BCSratioXY}), the ratio
(\ref{BCSXXX}) does not depend on the microscopic parameters of the model, and
even the dimerization amplitude $A_\epsilon$ cancels. In this sense we
interpret this as a universal result. The interactions renormalize the BCS
ratio away from the free fermionic value of 1.76.
%
%

%
%
$\bullet$ {\em \underline{Conclusions.}} --The BCS ratio in the interacting
Ising-XXX model is calculated. Similar to the classical free-fermionic case,
this value is also universal in the sense that is does not depend on the
microscopic parameters of the model. We conjecture that it must be the same for
various types of the Peierls-like transitions where the spin gap is accompanied
by the structural, orbital or charge order. An extension of these results
taking into account marginal terms of the spin chain Hamiltonian is warranted.
%
%

%
%
$\bullet$ {\em \underline{Acknowledgements.}} --
I am very grateful to C. Gros, J. Leeson for the  earlier spin-SAF theory
collaborations, and to S.L. Lukyanov for numerous enlightening discussions.
The author  acknowledges financial support from
NSERC (Canada) and the Laurentian University Research Fund.
%
%


\end{document}